\documentclass{article}
\usepackage[T1]{fontenc} % add special characters (e.g., umlaute)
\usepackage[utf8]{inputenc} % set utf-8 as default input encoding
\usepackage{amsmath,cite,url}
\usepackage{graphicx}
\usepackage[lbd]{ismir}
\usepackage{color}

\usepackage{float} % add this to your preamble

\usepackage[capitalize]{cleveref}

\usepackage[firstpage]{draftwatermark}
\definecolor{lightgray}{rgb}{0.9,0.9,0.9}
\definecolor{darkgray}{rgb}{0.4,0.4,0.4}
\SetWatermarkFontSize{12pt}
\SetWatermarkScale{1.1}
\SetWatermarkAngle{90}
\SetWatermarkHorCenter{202mm}
\SetWatermarkVerCenter{170mm}
\SetWatermarkColor{darkgray}
\SetWatermarkText{Late-Breaking / Demo Session Extended Abstract, ISMIR 2025 Conference}

\title{MaskBeat: Loopable Drum Beat Generation}

\multauthor
{Luca A. Lanzendörfer \hspace{0.5cm} Florian Grötschla \hspace{0.5cm} Karim Galal \hspace{0.5cm} \textbf{Roger Wattenhofer}}
{
ETH Zurich\\
{\tt\small \{lanzendoerfer, fgroetschla, kgalal, wattenhofer\}@ethz.ch}
}

\def\authorname{L. A. Lanzendörfer, F. Grötschla, K. Galal, and R. Wattenhofer}

\begin{document}

\maketitle

\begin{abstract}
We present MaskBeat, a transformer-based approach for loopable drum pattern generation. Rather than predicting drum hits sequentially, our method uses bidirectional attention with iterative refinement, allowing instruments to be generated in parallel while maintaining musical coherence. Additionally, we introduce custom loss functions that capture drum-specific musical relationships. Our experiments show that MaskBeat generates higher quality and more musically coherent drum patterns than baseline approaches.
\end{abstract}

\section{Introduction}
Generating realistic drum patterns presents several interesting challenges that distinguish it from other music generation tasks. Drums involve complex relationships between instruments, for instance, kick and snare drums typically follow complementary patterns, while closed and open hi-hats are physically mutually exclusive. Different musical genres also have distinct rhythmic signatures, and effective drum patterns need to maintain a consistent groove while still providing musical variation while still providing musical variation \cite{gillick2019learning,simon2018generating,liang2019midisandwich2rnnbasedhierarchicalmultimodal,roberts2018hierarchical,wei2019structured_vae_drum,nistal2020drumgan,lavault2022stylewavegan,aouameur2019neural_drum_machine}

Most existing approaches use autoregressive transformers that predict drum hits one instrument at a time \cite{huang2018music,oore2018time,jukedrummer, drumfillgen,jajoria2024textconditionedsymbolicdrumbeat,shepardson2022notochord}. However, we observed that this sequential prediction creates problems: Each decision only considers past context, which makes it difficult to maintain global groove consistency. The resulting patterns often suffer from temporal drift and cannot easily be edited or refined.

Our approach, MaskBeat, takes inspiration from MaskGIT \cite{chang2022maskgit} and reframes drum generation as a masked token prediction problem. The key insight is that drum patterns benefit from bidirectional context, knowing what happens in future beats can inform better decisions about the current beat. We make several contributions: (1) adapting MaskGIT to drum loops via timestep and instrument masking, (2) introducing groove and dependency losses that embed drum-specific musical knowledge, and (3) running ablation studies showing that MaskBeat surpasses baselines in musical quality and controllability.

A demo and the codebase of MaskBeat are available at \url{https://maskbeat.ethz.ch}

\section{Methodology}

% \subsection{Architecture and Tokenization}
\textbf{Token Representation.} We represent each timestep as a 9-dimensional binary vector $\mathbf{x}_t \in \{0,1\}^9$ that encodes which drums are active: [Kick, Snare, Closed Hi-Hat, Open Hi-Hat, Low Tom, Mid Tom, High Tom, Crash, Ride]. Our patterns span 32 steps (2 measures) at 16th-note resolution. This differs from previous work \cite{dong2017musegan} which treats instruments independently.

\textbf{MaskBeat Architecture.} Built on the MaskGIT backbone, our 8-layer, 8-head transformer ($d_{model}=512$) includes three drum-specific tweaks: (1) \emph{Dual Embedding}: a learned vector for the mask token (id 10) and a linear projection for each 9-D timestep; (2) \emph{Alternating masking}: training begins with full-timestep masks to learn groove, then gradually mixes in per-instrument masks for finer detail; (3) \emph{Fixed 32-step timing signal}: we add a simple repeating sine/cosine pattern so the network can tell "early beat" from "late beat" within every two-bar loop.

% \subsection{Musical Loss Functions}
\textbf{Instrument Dependency Loss (DL).} In addition to binary cross-entropy, we introduce novel losses that encode musical domain knowledge. Human drummers follow certain musical conventions that our model needs to learn. For example, kick drums typically appear on strong beats (beats 1 and 3), snare drums emphasize backbeats, and it is physically impossible to play both closed and open hi-hat simultaneously. Our dependency loss enforces these relationships: $\mathcal{L}_{dep} = \lambda_{ks}\!\sum_{i} \big[(1-\hat{y}_{i,\text{kick}}) + (1-\hat{y}_{i+4,\text{snare}})\big] + \lambda_{hh}\!\sum_{t} \hat{y}_{t,\text{CHH}}\, \hat{y}_{t,\text{OHH}} + \lambda_{tom}\,\mathcal{L}_{\text{tom}}$\,, where $\hat{y}_{t,\text{instr}}$ denotes the model's sigmoid output for a given instrument at timestep $t$, $\lambda_{ks}{=}0.15$, $\lambda_{hh}{=}0.3$ and $\lambda_{tom}{=}0.1$ weight the kick-snare, hi-hat and tom terms, respectively.

% Without this loss, the model might generate musically nonsensical patterns like constant snare hits or simultaneous open/closed hi-hats.

\textbf{Groove Loss (GL).} One key challenge in drum generation is maintaining temporal coherence. Patterns should feel like they have a consistent "pocket" or groove rather than random hits. Our groove loss addresses this by: (1) emphasizing musically important beat positions, and (2) encouraging consistency between similar timepoints across measures: $\mathcal{L}_{groove} = \sum_{t} w_t\, \mathcal{L}_{\text{sub}}(t) + \beta \sum_{m} \lVert\mathbf{P}_{m} - \mathbf{P}_{m+1}\rVert^2$\,, where $w_t\in\{1,2,4\}$ emphasises beat importance, $\mathcal{L}_{\text{sub}}(t)$ is the squared difference between activations at $t$ and $t{-}1$, $\mathbf{P}_m$ is the pattern in bar $m$, and $\beta{=}0.3$ controls the inter-bar consistency term.

% This helps prevent the "random noise" patterns that often emerge from purely data-driven approaches and instead produces patterns with clear rhythmic structure.

\textbf{Focal Loss (FL).} Drum patterns are naturally sparse, most timestep/instrument combinations are silent (0), with only a few active hits (1). Standard binary cross-entropy treats all predictions equally, but this means the model can achieve low loss by simply predicting mostly zeros everywhere. We counteract this by using a focal loss $\mathcal{L}_{focal} = -\alpha_t (1-p_t)^{\gamma} \log p_t$\,, where $p_t$ is the predicted probability for the true label at timestep $t$, $\alpha_t$ balances classes and we set $\gamma{=}2.0$ following \cite{lin2017focal}.

\section{Evaluation \& Application}

% \subsection{Dataset Construction and Preprocessing}
We construct a dataset from 30k professional drum loops spanning diverse musical genres. We use Groove Monkee\footnote{\url{https://groovemonkee.com/}} which contains 18k loops covering rock, jazz, latin, funk, and world music styles as well as Lakh MIDI~\cite{raffel2016learning} with 12k drum tracks from multi-instrument MIDI files. Our dataset covers tempo ranges 60-180 BPM with balanced genre distribution: rock (28\%), electronic (22\%), funk (18\%), jazz (15\%), world (17\%)\cite{hsiao2021compound}.

\textbf{Preprocessing Dataset.} We run various stages of preprocessing to ensure high-quality: (1) Swing-preserving quantization, maintains groove feel while normalizing to 16th-note grid using adaptive quantization that preserves swing ratios. (2) Velocity-aware processing, converts MIDI velocities to binary activations using genre-specific thresholds learned from professional recordings~\cite{gigamidi}. (3) Pattern deduplication, removes near-duplicates using Jaccard similarity with threshold 0.85. (4) Quality filtering, eliminates patterns with excessive density (>40\% active cells) or minimal content (<5\% active cells). (5) Augmentation strategy, time-shifting (1-16 steps), density adjustment (±20\%), and tempo-preserving stretching.

\textbf{Loop Extraction.} To prepare training data from multi-instrument MIDI files we first isolate the drum track (track with \texttt{is\_drum}=True) and quantize it to a 16th-note grid. We then down-mix the resulting 9\,$\times$\,$T$ roll to a 1-D activation vector $a\in\{0,1\}^T$ and compute its normalised autocorrelation $\varphi(\tau)=\tfrac{1}{T}\sum_t a_t a_{t+\tau}$. Searching for lags $\tau\in[16,64]$, we pick the shortest period $\tau^*$ with $\varphi(\tau^*)>0.8$, a strong indicator of repetition. The contiguous 32-step (two-bar) window starting at that phase is extracted as a loopable drum beat; segments that fail this criterion are discarded. This automatic procedure removes intros/outros and guarantees that every training example can be seamlessly repeated. After additional filters ($\geq 6$ drum hits, density $<40\%$) we obtain the 30k loops used for training.

% \textbf{Training Infrastructure and Optimization.} We use distributed training on 2 GPUs with gradient accumulation and mixed precision. AdamW~\cite{loshchilov2017decoupled} is used with $lr=10^{-4}$, $\beta_1=0.9$, $\beta_2=0.999$, weight decay $10^{-5}$, and batch size 24 per GPU (96 effective). Learning rate scheduling is a 15-epoch linear warmup followed by cosine annealing. Training converges after 150 epochs (ca.~8 hours). For regularization we 10\% dropout, 0.05 label smoothing, and gradient clipping (max norm=1.0).

% \textbf{Training Infrastructure and Optimization.} We use AdamW~\cite{loshchilov2017decoupled} with $lr=10^{-4}$, weight decay $10^{-5}$, and default $\beta$ parameters. Learning rate scheduling is a linear warmup followed by cosine annealing. Training converges after 150 epochs (ca.~8 hours).
% Additionally, we use 10\% dropout, 0.05 label smoothing, and gradient clipping (max norm=1.0).

% \subsection{Baselines and Metrics}
% We run an ablation against (1) AR Transformer (sequential drum generation), (2) MaskGIT, (3) MaskGit + GL, (4) MaskGit + DL, (5) MaskGit + Both Losses, and (6) MaskBeat (Full).

We evaluate different baselines to demonstrate the effectiveness of our proposed losses. As baselines we use an autoregressive transformer~\cite{vaswani2017attention}, MaskGit~\cite{chang2022maskgit}, and MaskGit using our losses. We evaluate using the following metrics: \textbf{Beat strength}, which quantifies adherence to musical metrical hierarchy by measuring the ratio of activations on strong beats (1, 3) versus weak beats (2, 4), where higher values indicate more musically structured patterns. \textbf{Pattern repetition}, measuring cosine similarity between consecutive 2-measure segments, measuring groove consistency essential for danceable rhythms (higher means better groove maintenance). \textbf{Instrument balance}, measuring Shannon entropy across 9-instrument activation frequencies, capturing rhythmic diversity without bias toward specific drums (higher means more balanced orchestration).
\cref{tab:results} demonstrates the performance of each model, indicating that the losses introduced in MaskBeat improve model performance.

\textbf{Novelty.}  To ensure the model is not memorizing and reproducing the dataset, we compare 1k generated loops against the full training set using token‐level intersection-over-union (IoU).  No sample exceeded an IoU of 0.90 (exact copy), and the median similarity was 0.31, confirming that MaskBeat produces novel patterns.

\begin{table}[t]
\begin{center}
\begin{small}
\begin{tabular}{lccc}
\hline
\textbf{Method} & \textbf{Beat Str.} & \textbf{Pat. Rep.} & \textbf{Instr. Bal.} \\
\hline
Autoregressive & 0.723 & 0.534 & 1.87 \\
MaskGIT (MG) & 0.758 & 0.578 & 2.01 \\
MG + GL & 0.789 & 0.645 & 2.08 \\
MG + DL & 0.821 & 0.621 & 2.18 \\
MG + GL + DL & 0.834 & 0.673 & 2.25 \\
\textbf{MaskBeat (ours)} & \textbf{0.847} & \textbf{0.692} & \textbf{2.31} \\
\hline
\end{tabular}
\end{small}
\end{center}
\caption{Ablation study demonstrating the usefulness of the introduced losses. For all metrics, higher means better.}
\label{tab:results}
\end{table}

\begin{figure}[t]
  \centering
  \includegraphics[width=\columnwidth]{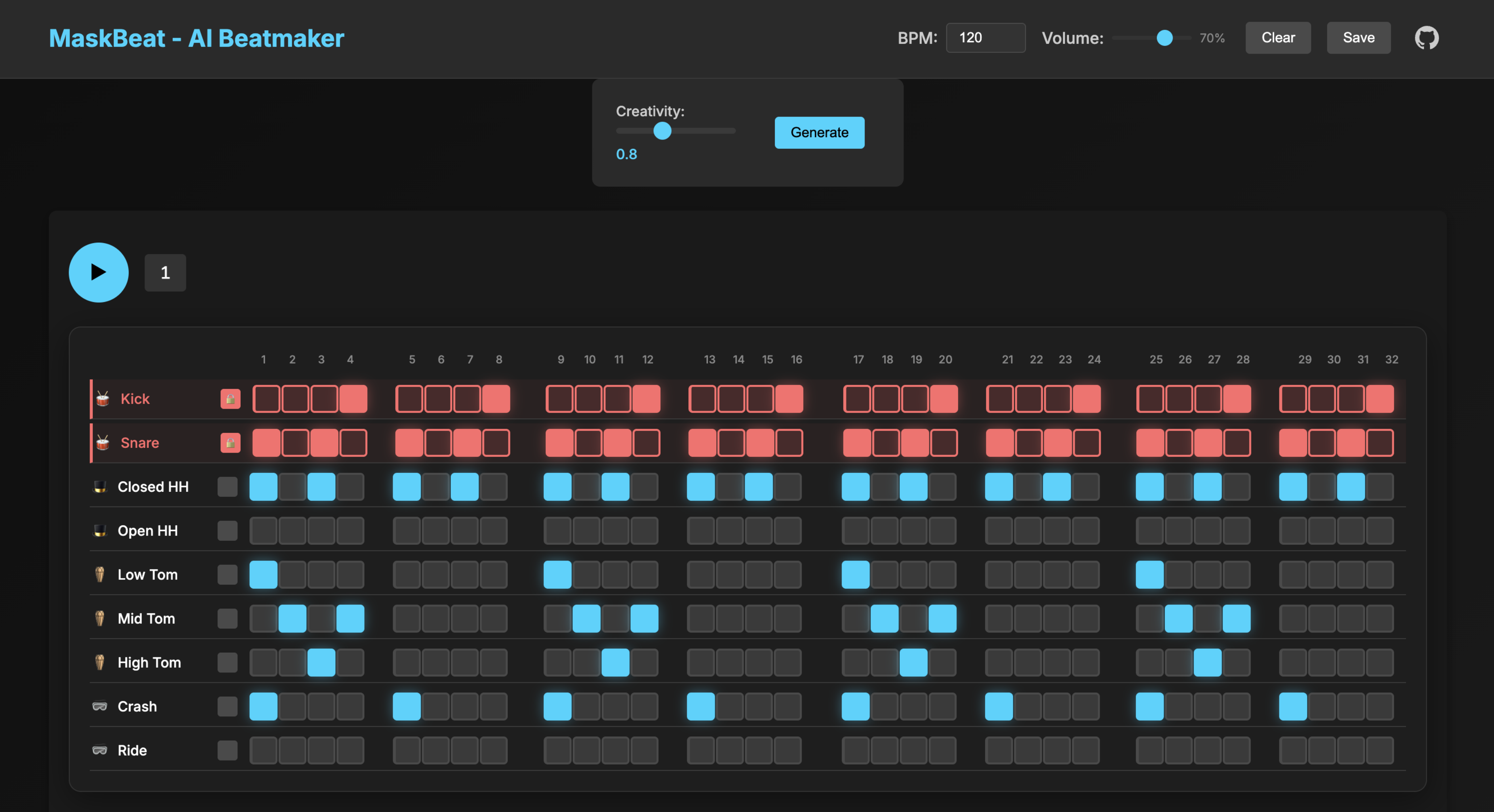}
  \caption{MaskBeat web application interface showing a generated drum pattern. Users can lock elements (shown in red), which are not altered during generation.}
  \label{fig:webapp_locked}
\end{figure}

% \section{Interactive Web Application}

\textbf{Web App.} We developed an interactive application (cf.~\cref{fig:webapp_locked}). The interface 
% was inspired by Splice Beatmaker\footnote{\url{https://splice.com/sounds/beatmaker}} and 
presents the user with a 9×32 grid representing the drum pattern, where each row corresponds to a drum instrument, and each column represents a 16th-note timestep. Users can lock individual instruments by clicking on instrument rows, which are then not altered during inference.
% \cite{eghbal-zadeh2018gan_drum_ui,vogl2019automatic_touch_ui_drum}
Furthermore, the user can control the sampling temperature using the ``creativity'' slider. Lower values make the softmax sharper, producing conservative, corpus-like grooves; higher values flatten the distribution and yield riskier, more inventive rhythms.

In summary, we believe MaskBeat represents a step in controllable drum generation by addressing a fundamental challenge: how to create patterns that are both musically coherent and creatively interesting. By incorporating knowledge about drum relationships and groove structure, we achieved significant improvements across all metrics.
% across our metrics: +17\% beat strength, +30\% pattern repetition, +24\% instrument balance.

\bibliography{ISMIRtemplate}

\begin{thebibliography}{10}
\providecommand{\url}[1]{#1}
\csname url@samestyle\endcsname
\providecommand{\newblock}{\relax}
\providecommand{\bibinfo}[2]{#2}
\providecommand{\BIBentrySTDinterwordspacing}{\spaceskip=0pt\relax}
\providecommand{\BIBentryALTinterwordstretchfactor}{4}
\providecommand{\BIBentryALTinterwordspacing}{\spaceskip=\fontdimen2\font plus
\BIBentryALTinterwordstretchfactor\fontdimen3\font minus \fontdimen4\font\relax}
\providecommand{\BIBforeignlanguage}[2]{{%
\expandafter\ifx\csname l@#1\endcsname\relax
\typeout{** WARNING: IEEEtran.bst: No hyphenation pattern has been}%
\typeout{** loaded for the language `#1'. Using the pattern for}%
\typeout{** the default language instead.}%
\else
\language=\csname l@#1\endcsname
\fi
#2}}
\providecommand{\BIBdecl}{\relax}
\BIBdecl

\bibitem{gillick2019learning}
J.~Gillick, A.~Roberts, J.~Engel, D.~Eck, and D.~Bamman, ``Learning to groove with inverse reinforcement learning,'' in \emph{Advances in Neural Information Processing Systems}, vol.~32, 2019.

\bibitem{simon2018generating}
I.~Simon, A.~Sarroff, and A.~Roberts, ``Learning a latent space of multitrack measures,'' 2018.

\bibitem{liang2019midisandwich2rnnbasedhierarchicalmultimodal}
\BIBentryALTinterwordspacing
X.~Liang, J.~Wu, and J.~Cao, ``Midi-sandwich2: Rnn-based hierarchical multi-modal fusion generation vae networks for multi-track symbolic music generation,'' 2019. [Online]. Available: \url{https://arxiv.org/abs/1909.03522}
\BIBentrySTDinterwordspacing

\bibitem{roberts2018hierarchical}
A.~Roberts, J.~Engel, C.~Raffel, C.~Hawthorne, and D.~Eck, ``A hierarchical latent vector model for learning long-term structure in music,'' in \emph{International Conference on Machine Learning}, 2018.

\bibitem{wei2019structured_vae_drum}
I.~Wei, C.~Wu, and L.~Su, ``Generating structured drum patterns using variational autoencoder and self‑similarity matrix,'' in \emph{Proceedings of ISMIR}, 2019, pp. 847--854, symbolic drum‑pattern generation with VAE + structural constraints.

\bibitem{nistal2020drumgan}
J.~Nistal, S.~Lattner, and G.~Richard, ``Drumgan: Synthesis of drum sounds with timbral feature conditioning using generative adversarial networks,'' \emph{arXiv preprint arXiv:2008.12073}, 2020, audio‑domain drum sound generator, not symbolic pattern.

\bibitem{lavault2022stylewavegan}
A.~Lavault, A.~Roebel, and M.~Voiry, ``Stylewavegan: Style‑based synthesis of drum sounds with extensive controls using generative adversarial networks,'' \emph{arXiv preprint arXiv:2204.00907}, 2022, audio drum‑sound generation, conditioned on drum type and descriptors.

\bibitem{aouameur2019neural_drum_machine}
C.~Aouameur, P.~Esling, and G.~Hadjeres, ``Neural drum machine: An interactive system for real‑time synthesis of drum sounds,'' \emph{arXiv preprint arXiv:1907.02637}, 2019, max4Live interface for real‑time drum‑sound synthesis.

\bibitem{huang2018music}
C.-Z.~A. Huang, A.~Vaswani, J.~Uszkoreit, N.~Shazeer, I.~Simon, C.~Hawthorne, A.~Dai, M.~Hoffman, M.~Dinculescu, and D.~Eck, ``Music transformer: Generating music with long-term structure,'' in \emph{International Conference on Machine Learning}, 2019, pp. 1364--1372.

\bibitem{oore2018time}
S.~Oore, I.~Simon, S.~Dieleman, D.~Eck, and K.~Simonyan, ``This time with feeling: Learning expressive musical performance,'' \emph{Neural Computing and Applications}, vol.~32, no.~4, pp. 955--967, 2020.

\bibitem{jukedrummer}
\BIBentryALTinterwordspacing
A.~Wang \emph{et~al.}, ``Jukedrummer: Conditional beat-aware audio-domain drum accompaniment generation via transformer vq-vae,'' \emph{ArXiv}, 2023, taiwan AI Labs. [Online]. Available: \url{https://arxiv.org/abs/2305.12644}
\BIBentrySTDinterwordspacing

\bibitem{drumfillgen}
\BIBentryALTinterwordspacing
H.~Shah \emph{et~al.}, ``Generating coherent drum accompaniment with fills and improvisations,'' \emph{arXiv preprint arXiv:2209.00291}, 2022. [Online]. Available: \url{https://arxiv.org/pdf/2209.00291}
\BIBentrySTDinterwordspacing

\bibitem{jajoria2024textconditionedsymbolicdrumbeat}
\BIBentryALTinterwordspacing
P.~Jajoria and J.~McDermott, ``Text conditioned symbolic drumbeat generation using latent diffusion models,'' 2024. [Online]. Available: \url{https://arxiv.org/abs/2408.02711}
\BIBentrySTDinterwordspacing

\bibitem{shepardson2022notochord}
V.~Shepardson, J.~Armitage, and T.~Magnusson, ``Notochord: a flexible probabilistic model for embodied midi performance,'' in \emph{Proceedings of the 3rd Conference on AI Music Creativity (AIMC 2022)}, Sep. 2022.

\bibitem{chang2022maskgit}
H.~Chang, H.~Zhang, L.~Jiang, C.~Liu, and W.~T. Freeman, ``Maskgit: Masked generative image transformer,'' in \emph{Proceedings of the IEEE/CVF Conference on Computer Vision and Pattern Recognition}, 2022, pp. 11\,315--11\,325.

\bibitem{dong2017musegan}
H.-W. Dong, W.-Y. Hsiao, L.-C. Yang, and Y.-H. Yang, ``Musegan: Multi-track sequential generative adversarial networks for symbolic music generation and accompaniment,'' in \emph{AAAI Conference on Artificial Intelligence}, 2018.

\bibitem{lin2017focal}
T.-Y. Lin, P.~Goyal, R.~Girshick, K.~He, and P.~Dollár, ``Focal loss for dense object detection,'' in \emph{Proceedings of the IEEE International Conference on Computer Vision}, 2017, pp. 2980--2988.

\bibitem{raffel2016learning}
C.~Raffel, \emph{Learning-based methods for comparing sequences, with applications to audio-to-midi alignment and matching}.\hskip 1em plus 0.5em minus 0.4em\relax Columbia University, 2016.

\bibitem{hsiao2021compound}
W.-Y. Hsiao, H.-W. Dong, and Y.-H. Yang, ``Compound word transformer: Learning to compose full-song music over dynamic directed hypergraphs,'' in \emph{AAAI Conference on Artificial Intelligence}, 2021.

\bibitem{gigamidi}
\BIBentryALTinterwordspacing
I.~Simon \emph{et~al.}, ``The gigamidi dataset with features for expressive music performance detection,'' \emph{MetaCreation Research Project}, 2023. [Online]. Available: \url{https://www.metacreation.net/projects/gigamidi-dataset}
\BIBentrySTDinterwordspacing

\bibitem{vaswani2017attention}
A.~Vaswani, N.~Shazeer, N.~Parmar, J.~Uszkoreit, L.~Jones, A.~N. Gomez, L.~Kaiser, and I.~Polosukhin, ``Attention is all you need,'' \emph{Advances in Neural Information Processing Systems}, vol.~30, 2017.

\end{thebibliography}

\end{document}